\documentclass[aps,prx,twocolumn,superscriptaddress,longbibliography]{revtex4-1}
\usepackage[T1]{fontenc}
\usepackage[latin9]{inputenc}
\usepackage{graphicx,amssymb,amsmath,xcolor}
\usepackage[unicode=true,
 bookmarks=false,
 breaklinks=false,pdfborder={0 0 0},pdfborderstyle={},backref=false,colorlinks=true]
 {hyperref}
 \hypersetup{
 colorlinks=true,
 linkcolor=blue, 
 citecolor=blue,  
 urlcolor=blue }
 \usepackage[normalem]{ulem}
\usepackage{xcolor}


\begin{document}
\title{Electronic structure, magnetic interactions, and magnonics of 2D trichloride materials}
\author{Krishna Prasad Chapai}
\thanks{These authors contributed equally to this work.}
\affiliation{Central Department of Physics, Tribhuvan University, Kirtipur, 44613 Kathmandu, Nepal}
\affiliation{Graduate School of Science and Technology, Mid-West University, Surkhet, Nepal}
\author{Yogendra Limbu}
\thanks{These authors contributed equally to this work.}
\affiliation{Department of Physics and Astronomy, University of Iowa, Iowa City, Iowa 52242, USA}
\author{Gopi Chandra Kaphle}
\affiliation{Central Department of Physics, Tribhuvan University, Kirtipur, 44613 Kathmandu, Nepal}
\author{Durga Paudyal}
\email{durga.quantum@gmail.com}
\affiliation{Department of Physics and Astronomy, University of Iowa, Iowa City, Iowa 52242, USA}


\begin{abstract}

\noindent We report phase stability, electronic structure, magnetic interactions, phononics, and magnonics of selective pristine, \(3d\)-element (Ti or Cr) doped, and \(4f\)-element (Ce) doped transition metal trichlorides (\(\text{M}\text{Cl}_3\)), calculated using advanced density functional theory. Negative cohesive and formation energies prove that pristine \(\text{M}\text{Cl}_3\) has strong chemical and structural stability. Meanwhile, negative defect formation energies show that adding dopant atoms is energetically favorable and keeps the doped crystal stable. The positive phonon frequencies in \(\text{Ce}\text{Cl}_3\) confirm the dynamical stability of adding cerium to the transition metal trichlorides. The phonon dispersions in this material exhibits exotic topological feature at a special symmetry direction $K$. Generalized gradient approximation (GGA) fails to capture the semiconducting nature of \(\text{M}\text{Cl}_3\), incorrectly predicting metallic behavior. In contrast, hybrid functional calculations that includes 25\% of exact exchange yield band gaps in good agreement with available  literature: 2.54 eV (\(\text{VCl}_{3}\)), 4.01 eV (\(\text{CrCl}_{3}\)), and 2.61 eV (\(\text{TiCl}_{3}\)). We use these benchmarks to derive appropriate Hubbard \(U\) parameters for more efficient modeling of magnetic exchange interactions. The ferromagnetic (FM) nature of MCl\({}_{3}\) is evidenced by its positive magnetic exchange interactions. The calculated exchange constants are subsequently used to obtain magnon dispersions by diagonalizing the spin Hamiltonian within the Holstein-Primakoff transformation. The two calculated magnon branches in CrCl\({}_{3}\), originating from the two magnetic sublattices, exhibit topological features along the high-symmetry $K$ direction. This topological magnonic feature at $K$ overlaps with the topological feature seen in phonons leading to magnon-phonon coupling. These dispersions are in an excellent agreement with the available experimental magnon dispersions of CrI$_3$. Doping VCl\({}_{3}\) with Ti or Cr modifies the bandgap (to 1.32 eV and 2.74 eV, respectively) while maintaining ferromagnetism. Ce-doping in CrCl\({}_{3}\) preserves the FM semiconducting behavior, producing a localized \(4f\) state and spin as well as orbital magnetic moments. Originating from this localized \(4f\) state below the Fermi level, nearest-neighbor Ce-Ce coupling yields a large exchange constant, suggesting possibility of a robust magnetic phase. Overall, the electronic and magnetic tunability of these materials through \(3d\)- and \(4f\)-electron doping in this family of 2D materials advances the magnonic and phononic functionalities making them ideal for spintronics.

\end{abstract}
\maketitle

\section{Introduction}

\noindent After the discovery of graphene in 2004~\cite{novoselov2005two, novoselov2004electric}, there has been a rapidly growing interest in the field of two dimensional (2D) materials due to their modifiable surfaces, excellent mechanical strength, and tunable electronic, magnetic, and quantum  properties~\cite{kamysbayev2020covalent,limbu2025magnetic} leading to energy and quantum functionalities. The 2D materials possess different electronic and magnetic properties than their bulk counterparts due to quantum confinement and weak dielectric-screening~\cite{mak2010atomically}. They exhibit high carrier mobility \cite{novoselov2005two, novoselov2004electric, geim2007rise}, direct-to-indirect band gap transitions~\cite{mak2010atomically}, and non-trivial topological states~\cite{choudhary2020computational, lee2021multiple, tang2017quantum}, making them promising for spintronics~\cite{wolf2001spintronics, vzutic2004spintronics, bader2010spintronics}, magnetic sensors, and magnonics ~\cite{kumar2014magnonic}. Despite the challenge of achieving long-range magnetism at elevated temperatures in 2D family~\cite{mermin1966absence}, chromium iodide (CrI$_3$) has been experimentally identified to exhibit long-range FM order due to its strong out-of-plane magnetic anisotropy~\cite{huang2017layer}. Defects, particularly doping, in this family of 2D materials significantly influence electronic and magnetic properties, as well as chemical stability~\cite{limbu2025magnetic,limbu2022electronic}, which are crucial for realizing the functionalities in defect-related optical and spintronic devices~\cite{amer2024defects}. Here, we systematically investigate the influence of $3d$-and $4f$-electrons doping in MCl$_3$, which has attracted considerable interest due to the unique properties arising from localized magnetism, thereby advancing the rapidly growing field of quantum information science (QIS).
 
Monolayer transition metal (TM) trihalides \(\text{M}\text{X}_3\), where M is the TM and X is a halogen) have attracted significant interest for spintronics owing to their robust magnetic ordering at finite temperatures ~\cite{lin1993dimensional, wang2020prospects}. CrI$_3$, extensively studied trihalide monolayer is a FM semiconductor with Curie temperature of 45~K~\cite{huang2017layer}, which is slightly lower than its bulk crystal 61~K~\cite{mcguire2015coupling}. However, vanadium trihalides (VX$_3$) have attracted further interest due to their higher magnetic transition temperatures as compared to CrX$_3$~\cite{tomar2019}. The bulk vanadium triiodide (VI$_3$) has been experimentally identified as a FM semiconductor with Curie temperature of 49~K and band gap of 0.6~eV ~\cite{kong2019vi3, son2019bulk}. On the other hand, theoretical calculations report band gap of 1.26~eV in its monolayer form~\cite{tomar2019}. It also exhibits a large orbital magnetic moment, leading to a large magnetic anisotropy~\cite{hovancik2023large}. Further, photoemission experiments reveal it as a strongly correlated Mott-insulator~\cite{bergner2022polarization,de2022influence}. Within the VX$_3$ family, VCl$_3$ has been theoretically predicted to be a Dirac half-metal (DHM)~\cite{he2016unusual, sun2020intrinsic, zhou2016evidencing, ouettar2023tuned,zhang2024design}, however, experiments confirm Mott-insulating behavior~\cite{mastrippolito2023polaronic}. Similarly, CrCl$_3$ is reported as a FM semiconductor~\cite{zhang2015robust}, whereas TiCl$_3$ is reported as a FM DHM~\cite{zhou2016evidencing} as well as a semiconductor~\cite{geng2020magnetic}. The inconsistencies in these results may arise from the inadequate treatment of strong electron correlation, necessitating either the use of hybrid functional (HSE06) approach or the determination of an appropriate Hubbard $U$ parameter~\cite{bousquet2010j,ryee2018comparative,malyi2020false,perdew2021interpretations,gunnarsson1976exchange} for the GGA + $U$ method.

Magnons, commonly known as quantized quasi-particles representing collective spin excitations or spin waves in magnetically ordered materials, have attracted considerable interest because of their ability to transmit and process information with significantly lower energy dissipation than conventional charge-based electronics, making them promising information carriers for next-generation spintronic and quantum technologies~\cite{chumak2015magnon,pirro2021advances}. Experimentally, magnonic properties have been observed in layered magnetic materials such as CrBr$_3$~\cite{yelon1971renormalization} and CrI$_3$~\cite{chen2018topological}. Theoretical studies have predicted intriguing magnonics in 2D materials~\cite{limbu2025magnetic}. Studying magnons in 2D magnetic materials is a new and exciting field. Researchers still have much to learn about how spin waves move in these thin layers. This gap in our knowledge drives our current research. 

The electronic and magnetic properties of these trichlorides can be modulated by transition metal (TM) and rare-earth (RE) doping. Previously, $3d$-transition-metal doping in VCl$_3$ was reported, where Sc-, Ti-, and Cr-doping preserved the FM semiconductor state, whereas Mn- and Fe-doping drove the FM semiconductor into a spin-gapless FM semiconductor~\cite{ouettar2023tuned}. In contrast, ferrimagnetic half-metal has been predicted in Ni-doped CrCl$_3$~\cite{wang2019theoretical}. In addition to the TM, RE elements exhibit a unique electronic and magnetic properties  characterized by their strongly localized $4f$-electrons~\cite{benelli2002magnetism}. Particularly, Cerium (Ce) has only a single $4f$ electron, which produces a unique quantum state below the Fermi level~\cite{bhandari2023giant}. The RE  (e.g., Ce) doped wide band gap materials show a narrow and controllable linewidth,  allowing the $4f$ -- $4f$ optical transitions with long spin and optical coherence times, thereby advancing the QIS~\cite{limbu2025stability,limbu2025ab}.  

Here, we have systematically studied the phase stability, defect formation, electronic structure, and magnetic properties as well as magnon and phonon dynamics of selective pristine and doped (TMs and Ce) VCl$_3$, CrCl$_3$, and  TiCl$_3$ using density functional theory (DFT) calculations. The negative values of the cohesive and formation energies confirm the structural and chemical stability of the pristine materials, while the negative values of the defect formation energies show the chemical stability of the doped materials. The positive phonon frequencies in cerium trichloride confirm the dynamical stability of adding cerium to the TM trichlorides. The phonon dispersions in this material exhibits exotic topological feature at a special symmetry direction $K$. The electronic structure calculations with GGA underestimate the band gap of CrCl$_3$ and TiCl$_3$, and also predicts VCl$_3$ to be Dirac half metal in contrast with experimental result. The underestimated band gaps in CrCl$_3$ and TiCl$_3$ and half metallic nature of VCl$_3$ are then corrected using Hubbard $U$ on top of GGA, making them FM semiconductors, which are also confirmed from HSE06 calculations. The positive values of magnetic exchange interaction constants reveal the FM nature of all the materials. These calculated exchange interaction constants are used to obtain spin wave excitations, yielding two magnon branches associated with the two magnetic sublattices, and exhibiting topological features along the high-symmetry $K$ direction. This topological magnonic feature at $K$ correlates with the topological feature seen in phonons leading to magnon-phonon coupling.  Transition metals and Ce doped materials remain FM semiconductors. In particular, Ce doping in CrCl$_3$ becomes robustly FM by enhancing the magnetic exchange interactions and generates a localized $f_{y(3x^2-y^2)}$ state just below the Fermi level, yielding a spin magnetic moment of $\sim$~1~$\mu_B$ per Ce.
 
\section{Theoretical and computational approaches} 
Spin-polarized first-principles calculations were performed using the DFT~\cite{koch2015chemist}, with Vienna \textit{Ab initio} Simulation Package (VASP) \cite{hafner2008ab} within projector augmented-wave (PAW) method~\cite{blochl1994projector}. The structural and magnetic properties of these compounds are calculated using  Perdew-Burke-Ernzerhof (PBE) within generalized gradient approximation (GGA) \cite{perdew1996generalized}. We used VASP with the HSE06 hybrid functional~\cite{heyd2004efficient}, which accurately treats localized \(3d\)- and \(4f\)-electrons to determine reliable band gaps and magnetic moments. A plane-wave basis set with a kinetic energy cutoff of 500 eV and a \(8 \times 8 \times 1\) Monkhorst-Pack ~\cite{monkhorst1976special} \(k\)-point mesh were used for the calculations. The convergence thresholds for the total energy and forces were set to \(10^{-6}\) eV and \(10^{-5}\) eV/\text{\AA}, respectively. We also performed DFT calculations using Quantum Espresso (QE)~\cite{giannozzi2017advanced, giannozzi2020quantum} with the same exchange correlation functional and convergence parameters as used in VASP calculations. We used ultrasoft pseudo-potentials~\cite{vanderbilt1990soft} for V, Cr, Ti, and Cl atoms, respectively and PAW pseudo-potential for Ce with $5d^14f^16s^2$ valence electrons. We set a vacuum of 20~\AA~ to avoid the interactions between the  monolayers. Atomic positions and lattice parameters are optimized employing GGA. Strongly correlated \(3d\) and \(4f\) electrons are also modeled using the \(\text{GGA}+U\) method, in addition to the HSE06 approach, with spin-orbit coupling (SOC). The effective Hubbard parameters (\(U_{\text{eff}}\)) for the \(\text{GGA}+U\) calculations are set to 5~eV for V, 3~eV for Cr~\cite{xue2019two}, 3~eV for Ti, and 5~eV for Ce~\cite{loschen2007first}. 

The dynamical stability of CeCl$_3$ is assessed using the finite-displacement approach~~\cite{chaput2019finite}, in which only symmetry-inequivalent atomic displacements are considered. The second-order force-constant matrix is subsequently constructed from the resulting forces and diagonalized using the \textsc{Phonopy}~\cite{togo2023first}, yielding the phonon modes.

For the magnetic exchange interactions, we construct a $2\times2\times1$ supercell containing 32 atoms and consider four distinct magnetic configurations: FM and three AFM configurations, namely AFM zigzag, AFM N\'eel, and AFM stripy (Fig.~\ref{exchange}). The total energies of these configurations are mapped onto the Heisenberg spin Hamiltonian to extract the magnetic exchange coupling constants. The resulting exchange parameters are subsequently used to calculate magnon dispersions by diagonalizing the spin Hamiltonian within the linear spin-wave approximation via the Holstein--Primakoff transformation~\cite{holstein1940field}, which maps the spin operators onto bosonic operators.
		
\section{RESULTS AND DISCUSSION}

\subsection{Pristine  trichloride materials}

\subsubsection{Phase stability}

\begin{figure}[ht]
\centering
\includegraphics[width=0.45\textwidth]{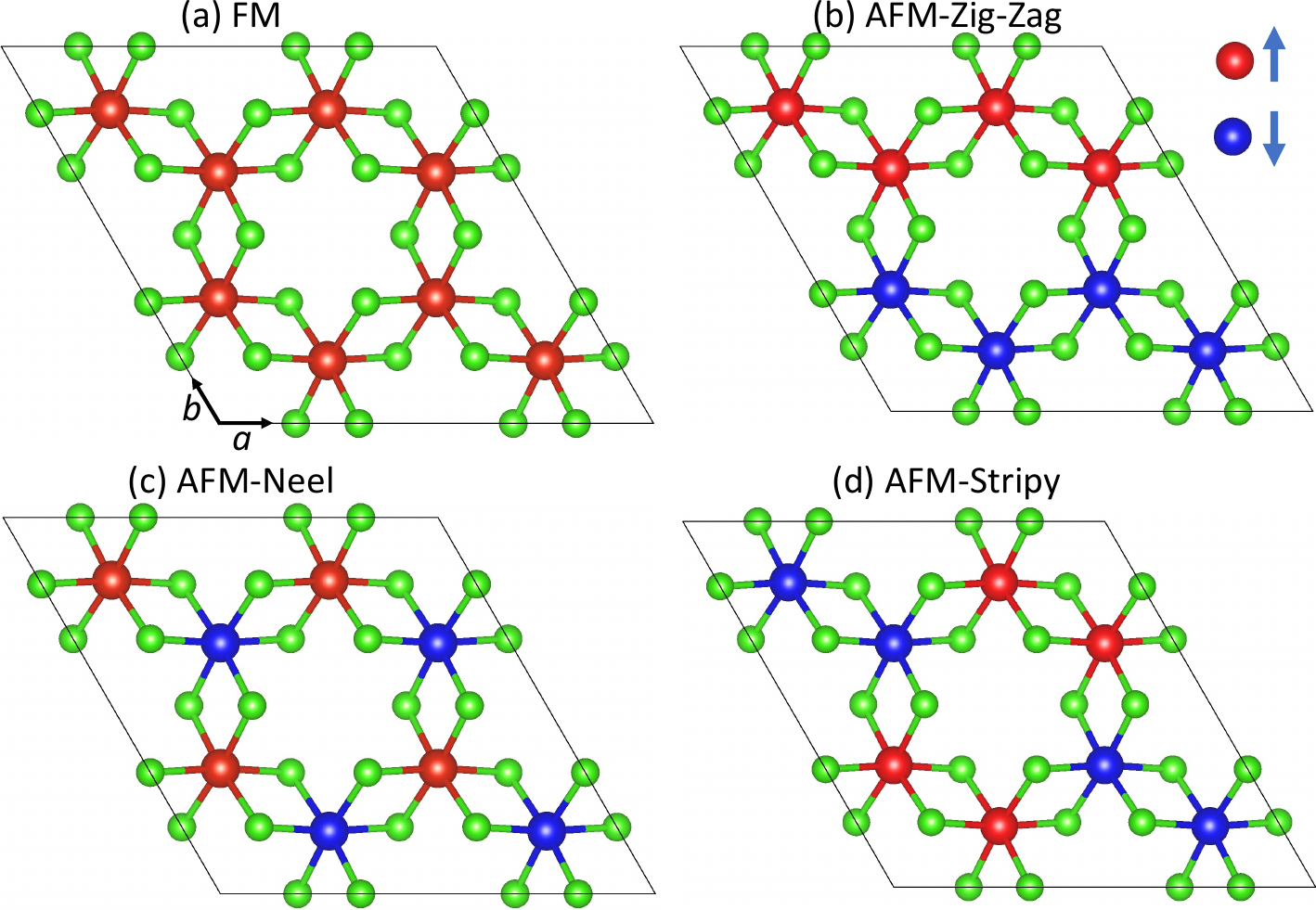}\hfill
\caption[dop]{ Top views of different magnetic configurations of VCl$_3$ in a 2 $\times$ 2 $\times$ 1 supercell: FM (a), AFM-Zig-Zag (b), AFM-N\'eel (c), and AFM-Stripy (d). Red (spin up) and blue (spin down) circles represent  V-atoms, while lime circle represents Cl-atom.}
\label{exchange}
\end{figure}

The monolayer TM trihalide forms with a crystal structure type of CrI$_3$ that belong to the $P\bar{3}1m$ space group, as also suggested in Ref.~\cite{liu2018screening}. This is the most energetically favorable structure. The vdW layered materials are cleaved into 2D materials~\citep{burch2018magnetism}.  The optimized structure of MCl$_3$ consists of Cl--M--Cl sandwich, where a sheet of M atoms is sandwiched between the two sheets of Cl atoms. The unit cell has two M and six Cl atoms, where each M atom is in octahedral coordination with six Cl atoms, forming a honeycomb network of M metal cations (Fig.~\ref{exchange}). The computed lattice constants, including bond lengths, for TiCl$_3$, VCl$_3$, and CrCl$_3$, which are presented in Table~\ref{crystal_information} are in good agreement with the available experimental values \citep{mccarley1964transport, klemm1947kristallstrukturen, morosin1964x}.

\begin{table}[b]
\caption{\label{tab:table1}%
{Calculated equilibrium lattice constants $a$, M--Cl bond length ($d_\text{M--Cl}$), M -- M bond length ($d_\text{M--M}$), and M--Cl--M bond angle in MCl$_3$ (M = Ti, V, and Cr).}}
\begin{ruledtabular}
    \begin{tabular}{l c c c c}
     Compounds & $a$ (\AA) & $d_{\text{M}-\text{Cl}}$ (\AA) & $d_{\text{M}-\text{M}}$ (\AA) & $\angle (\text{M}-\text{Cl}-\text{M})^\circ$ \\
     \hline 
     VCl$_3$  & $6.17 $      & $2.38$            & $3.56$            & $96.78$ \\
     CrCl$_3$ & $6.06$      & $2.36$            & $3.50$            & $95.47$ \\
     TCl$_3$ & $6.11$      & $2.42$            & $3.53$            & $93.25$ \\
    \end{tabular}
    \label{crystal_information}
\end{ruledtabular}
\end{table}

The structural and chemical stability of MCl$_3$ (M = V, Cr, and Ti) is checked using the  formula~\cite{limbu2025magnetic,limbu2022electronic},

\begin{equation}
E_\text{coh/for} = E_\text{total} - E^\text{M}_\text{iso/bulk} - 3 E^\text{Cl}_\text{iso/bulk}
\nonumber
\end{equation}

where $E_\text{total}$ is the total energy of the optimized material. $E^\text{M}_\text{iso}$ and $E^\text{M}_\text{bulk}$ represent the isolated and bulk energies of M (V, Cr, and Ti), while $E^\text{Cl}_\text{iso}$ and $E^\text{Cl}_\text{bulk}$ are the energies of the isolated and bulk Cl atom. The calculated cohesive energies ($E_\text{coh}$) are -3.73, -3.28, and -4.14~eV/atom, and the corresponding formation energies ($E_\text{for}$) are -1.27, -1.15, and -1.60~eV/atom for VCl$_3$, CrCl$_3$, and TiCl$_3$, respectively. The negative values of these cohesive and formation energies confirm their structural and chemical stability.

\subsubsection {Electronic band structure, density of states, band gaps, and magnetic moments} 

\begin{figure}
\centering
\includegraphics[width=0.30\textwidth]
{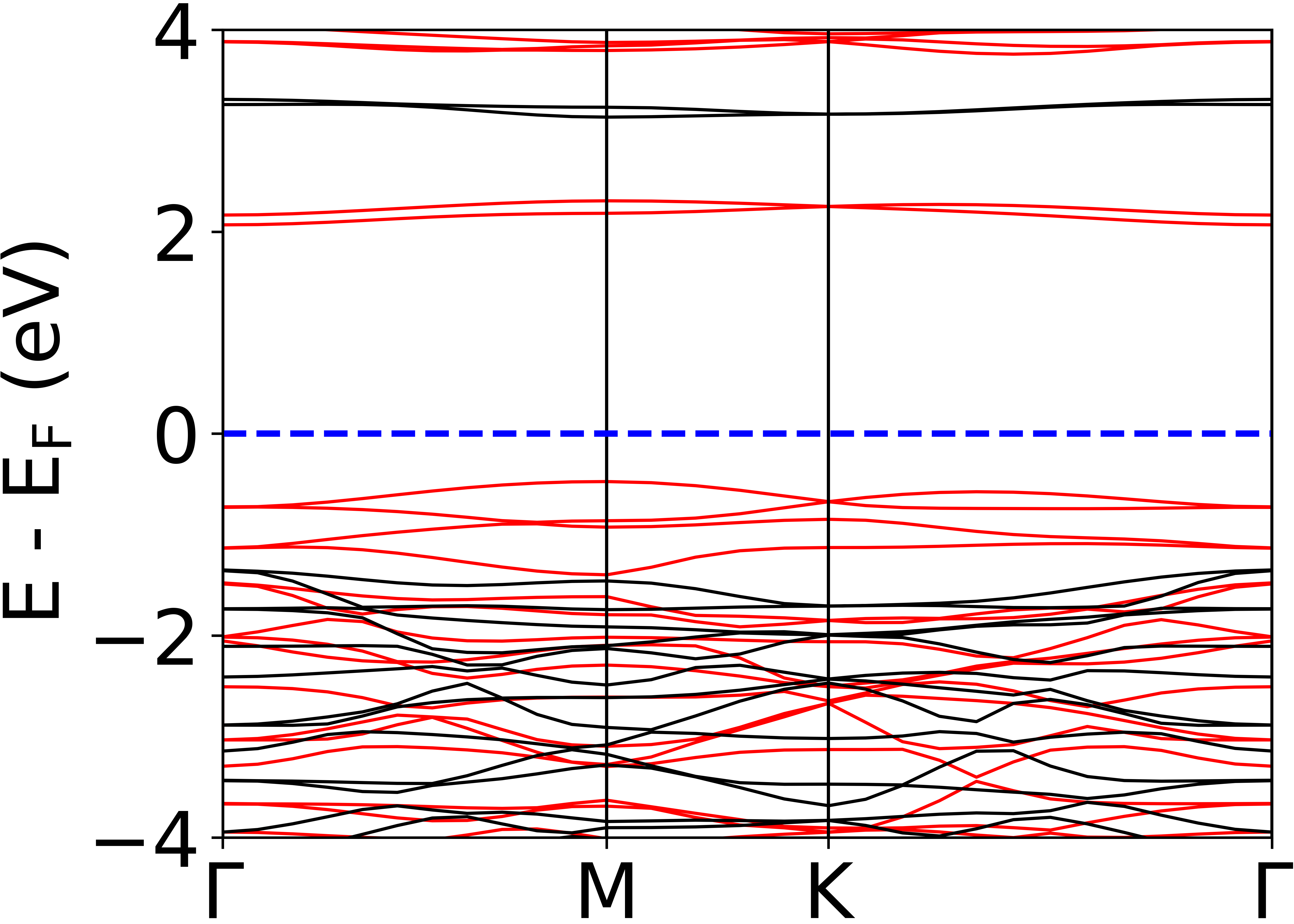}
\caption{Electronic band structure of pristine VCl$_3$ calculated using the HSE06.}
\label{fig:sidebyside}
\end{figure}

The electronic band structure of VCl$_3$ using GGA reveals it as Dirac half metal with no gap in the spin up channel and 3.71~eV gap in the spin down channel, which is in agreement  with the value reported in Ref.~\cite{he2016unusual}. By incorporating the Hubbard correction along with SOC effect, VCl$_3$ is identified as FM semiconductor with band gap of 2.41~eV, which is higher than the experimental and standard DFT value (1.8~eV)~\cite{mastrippolito2023polaronic}. The HSE06 calculations using VASP yield a band gap of 2.54~eV (Fig. \ref{fig:sidebyside}), comparable to the literature value (2.51~eV)~\cite{tomar2019}. We find semiconducting behavior in both CrCl$_3$ and TiCl$_3$ with band gaps of 2.45~eV and 2.19~eV with  DFT + $U$ ($U$ = 3~eV).  These are slightly smaller than the previously reported values~\cite{zhang2015robust,geng2020magnetic}. Interestingly, these are in contrast to the intrinsic half metallicity with half metallic gap of 0.6~eV as reported in previous studies using HSE06 functional calculations for TiCl$_3$~\cite{zhou2016evidencing}. Further, we performed HSE06 calculations for CrCl$_3$ and TiCl$_3$ and found them as semiconductors with band gaps of 4.01~eV and 2.61~eV, respectively. These are in good agreement  with previously reported values of 3.44~eV~\cite{zhang2015robust} and 3.84~eV~\cite{tomar2019} for CrCl$_3$ and 2.65~eV for TiCl$_3$~\cite{geng2020magnetic}. The electronic structure calculations and analysis performed and presented here with advanced functional like HSE06, therefore, pinpoints that these TM trichlorides are predominately FM semiconductors with reasonably wide band gaps critically important for static and excitonic functionalities. 

\begin{figure}
\centering
\includegraphics[width=0.49\textwidth]{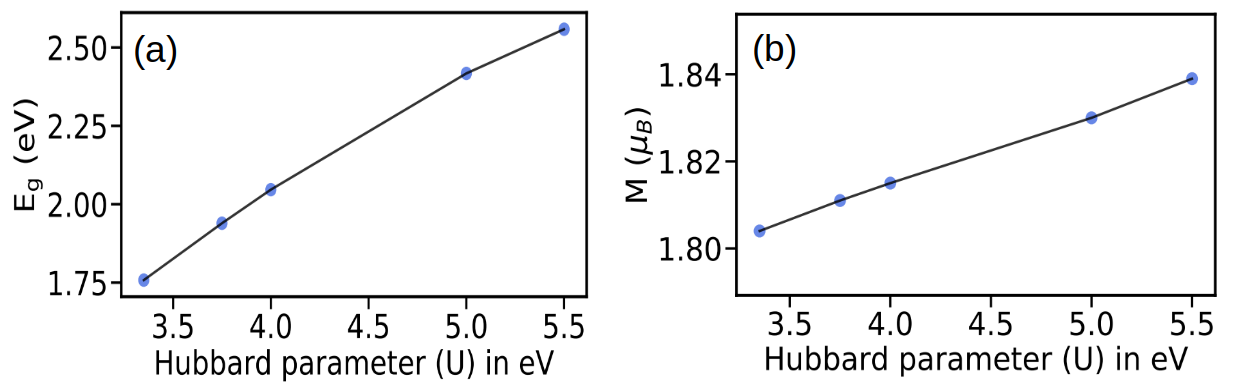}
\caption{Variation of band gap (a) and magnetic moment (b) as a function of the Hubbard parameter $U$ in VCl$_3$.}
\label{variation_with_U}
\end{figure}

In VCl$_3$, the crystal field generated by octahedrally coordinated chlorine atoms around V atom causes the splitting of $d$-orbitals into  $e_g$ and $t_{2g}$ levels. The degenerated $t_{2g}$ level splits into doubly degenerated  $a_{1g}$ and $e^\prime_g$ levels~\cite{de2022influence}. The insulating band gap is obtained due to the empty $a_{1g}$ states~\cite{nguyen2021electric}. The variation of band gap  and the atomic magnetic moment  of VCl$_3$ with the Hubbard $U$, including SOC is depicted in Fig.~\ref{variation_with_U}, revealing  almost linear with $U$ value. We find an inconsistency in the electronic band structure of VCl$_3$ between the PBE and PBE + $U$ + SOC/HSE06 levels, although the PBE results obtained from both QE and VASP remain consistent. In contrast, the electronic structures of CrCl$_3$ and TiCl$_3$ show consistent behavior across all computational approaches.

\begin{figure*}
\centering
\includegraphics[width=0.75\textwidth]{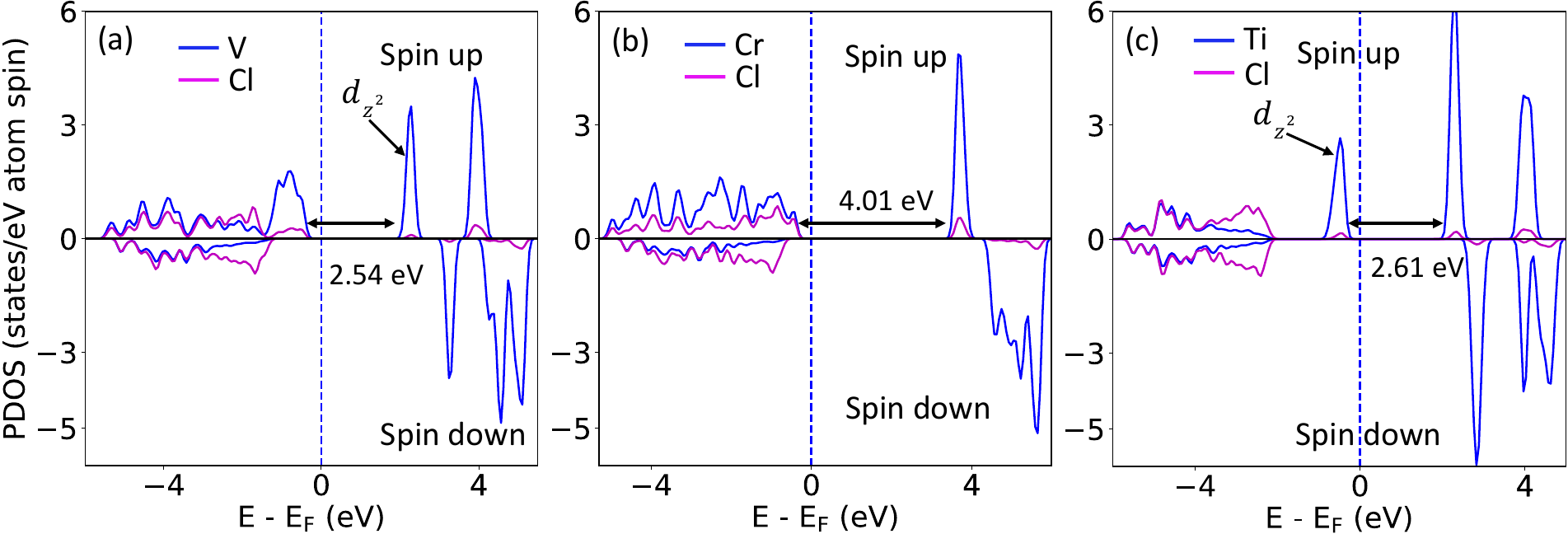}
\caption{Partial density of states (PDOS) per atom of pristine VCl$_3$ (a), CrCl$_3$ (b), and TiCl$_3$ (c) calculated using the HSE06. All the materials are found to be FM semiconductors. Interestingly, a single localized $3d$ ($d_{z^2}$) state is identified in TiCl$_3$, producing a spin magnetic moment of $\sim$ 1~$\mu_B$.}
\label{PDOS-hybrid}
\end{figure*}

The unit cell of VCl$_3$, CrCl$_3$, and TiCl$_3$ in FM states  have total magnetic moments of 4~$\mu_B$ (1.96~$\mu_B$/V), 6~$\mu_B$ (3.01~$\mu_B$/Cr), and 2~$\mu_B$ (0.98~$\mu_B$/Ti), respectively. The transition metals (TMs) V, Ti, and Cr contribute to the total magnetic moments, while the Cl atom contribute negligibly. The partial density of states (PDOS) of the pristine materials are calculated using HSE06 to study the atomic contribution to the electronic band structures (Fig. \ref{PDOS-hybrid}). The bottom of conduction band and top of valence band are contributed by TM-$3d$ states, which is responsible to produce the band gaps. There is a strong hybridization between TM-$3d$ and Cl-$2p$ states at the top of the valance band in both VCl$_3$ and CrCl$_3$ as shown in Fig.~\ref{PDOS-hybrid} (a and b). Interestingly, a single localized $d_{z^2}$ state is appeared just below the Fermi level, producing the spin magnetic moment~$\sim$ 1~$\mu_B$ in TiCl$_3$, while a single $d_{z^2}$ states is found at bottom of conduction band in VCl$_3$. 

\subsubsection{Magnetic exchange interactions}

To determine the magnetic ground state structure of VCl$_3$, we performed total energy calculations of the optimized 2 $\times$ 2 $\times$ 1 supercell of the FM and various antiferromagnetic (AFM) configurations, such as AFM-N\'eel, AFM-Zigzag, AFM-stripy (Fig.~\ref{exchange}). Our calculations show that the energy of VCl$_3$ in FM configuration is 264.49~meV lower than that of AFM-zigzag, 526.20~meV lower than AFM-ST, and 1022~meV than that of AFM-N\'eel configurations, confirming the FM ground state. The magnetic exchange interaction parameters $J_1$, $J_2$, and $J_3$, corresponding to the interaction between first-, second-, and third-nearest-neighbor magnetic ions, are extracted by mapping the total energies of the material with different magnetic configurations from the classical Heisenberg model:

\begin{equation}
\mathcal{H}_\text{spin} = -\sum_{i,j}J_1 \mathbf{S}_i \cdot\mathbf{S}_j - \sum_{k, l}J_2 \mathbf{S}_k \cdot\mathbf{S}_l -  \sum_{m, n}J_3 \mathbf{S}_m \cdot\mathbf{S}_n,
\label{Hamiltonian}
\nonumber
\end{equation}

Where S is the net magnetic moment at TM sites with $(i,j), (k, l)$  and $(m,n)$ representing the  first-, second-, and third-nearest-neighbor TM atoms, respectively. The value of $J_1$, $J_2$, and $J_3$ are then
obtained by mapping the total energies of different magnetic configurations into the Heisenberg  spin Hamiltonian from the following equations:\cite{sivadas2015magnetic,limbu2025magnetic}

\begin{equation}
\begin{aligned}
E_\text{FM/N\'eel} & = E_0 - (\pm 3J_1 + 6J_2 \pm 3J_3) \text{S}^2, \\
E_\text{ZZ/ST}  &= E_0 - (\pm J_1 - 2J_2 \mp 3J_3) \text{S}^2 
\end{aligned}
\nonumber
\end{equation}

The calculated values of $J_1$, $J_2$, and $J_3$ for VCl$_3$ are 30.09~meV, -6.77~meV, and 10.07~meV, respectively. The first- and third-nearest-neighbor exchange parameters between V atoms are positive, indicating FM coupling, whereas the second-nearest-neighbor interaction is negative, indicating AFM coupling between V atoms. The result shows that $J_1$ is one order of magnitude larger than $J_2$ and $J_3$ indicating the strong FM ground states due to the first neighboring V atoms. For CrCl$_3$, the exchange parameters are:  $J_1$ = 2.28~meV, $J_2$ = 0.24~meV, and $J_3$ = -0.17~meV, respectively, which are in good agreement with the values reported in Ref.~\cite{zhang2015robust}. Similarly, for TiCl$_3$, the corresponding exchange parameters are $J_1=35.730$ meV, $J_2=29.372$ meV, and $J_3=-27.817$ meV. The substantially larger first neighbor exchange constants in TiCl$_3$ and VCl$_3$ as compared to that of CrCl$_3$ is due to the drastic different electronic structure. There is a sharp $3d$ density of states peak just below the Fermi level in both TiCl$_3$ and VCl$_3$ indicating more localized magnetism arising from these states. For both CrCl$_3$ and TiCl$_3$, the first- and second-nearest-neighbor exchange interactions are positive, indicating FM coupling between the transition elements. In contrast, the third-nearest-neighbor exchange interactions are negative, indicating AFM coupling between the third nearest neighbor magnetic ions. This analysis, therefore, pinpoints the drastic changes in magnetic exchange interactions while going from Ti to V to Cr in trichloride.  

Further, the origin of exchange interactions involve the super-exchange interactions,  which drive the long range magnetic ordering  in the magnetic materials. According to Goodenough--Kanamori-Anderson~\cite{goodenough1955theory,kanamori1959superexchange,anderson1950antiferromagnetism}, the angle M--Cl--M is close to $90^0$  in super-exchange path for FM ordering in magnetic material  and it is $180^0$ for AFM ordering. The values of angle M--Cl--M for VCl$_3$, TiCl$_3$, and CrCl$_3$  are listed in Table ~\ref{crystal_information} and are close to $90^0$. This confirms that the dominant interaction is the super-exchange between two nearest-neighbor M atoms mediated by a Cl atom, leading to FM ground state. 

\subsubsection{Magnons}

\begin{figure}
\centering
\includegraphics[width=0.45\textwidth]{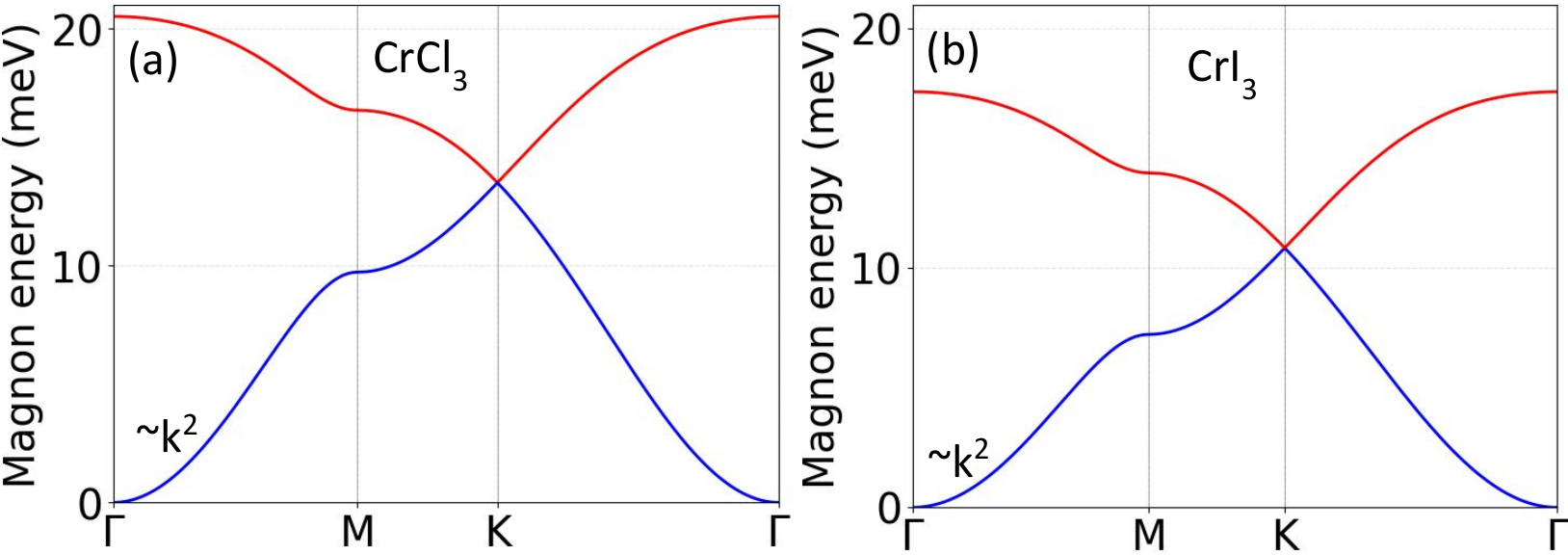}
\caption{Magnon dispersions of (a) CrCl$_3$, calculated using $J_1=2.28$ meV, $J_2=0.24$ meV, and $J_3=-0.17$ meV, and (b) CrI$_3$, calculated using the experimental exchange constants $J_1=2.01$ meV, $J_2=0.16$ meV, and $J_3=-0.08$ meV~\cite{chen2018topological}. Both materials exhibit an exotic topological feature along the high-symmetry K direction.}
\label{magnon}
\end{figure}

Using the exchange interaction constants obtained from DFT calculations, the spin Hamiltonian is diagonalized via the Holstein-Primakoff transformation~\cite{holstein1940field}. By mapping spin operators into bosonic operators, we obtain the magnetic excitations for FM CrCl$_3$. Details of the formalism is presented in Ref.~\cite{limbu2025magnetic}, and the same approach is adopted here. For CrCl$_3$, only the two exchange constants ($J_1$ and $J_2$) provide dominant contributions; thus, $J_3$ is neglected in the magnon dispersion calculations (Fig.~\ref{magnon}). There are two magnon bands in both materials, corresponding to the two Cr sublattices in the primitive cell. These magnon bands exhibit an approximately quadratic dependence on the wave vector $k$, as found in 2D ferromagnet~\cite{limbu2025magnetic}.  Interestingly, the two magnon branches cross at the high symmetry $K$ point with energy $\sim$ 13 meV, indicating a possible topological magnon feature. This magnonic topological feature correlates with the phononic topological feature, as discussed below in phonon section. Dzyaloshinskii-Moriya interactions (DMI) are not included in our model, which potentially opens a gap at the $K$ point. Surprisingly, the calculated magnon dispersion of CrCl$_3$ exhibits a similar dispersion behavior and comparable energy scale to the experimentally observed magnon spectrum of CrI$_3$~\cite{chen2018topological}, despite the difference in the halogen atoms. This is attributed to the comparable magnitudes of the underlying exchange interactions. In particular, the calculated exchange constants of CrCl$_3$ are of the same order as the experimentally determined values for CrI$_3$, $J_1=2.01$~meV, $J_2=0.16$~meV, and $J_3=-0.08$~meV~\cite{chen2018topological}. The resulting comparable exchange energy scales naturally lead to similar magnetic excitation spectra in these two iso-structural compounds (Fig.~\ref{magnon}). Experimentally, CrI$_3$ also exhibits a high-energy magnon mode at the zone center, and the magnon branches cross at the high symmetry $K$ point~\cite{chen2018topological}, as found in both CrCl$_3$ and VCl$_3$. In VCl$_3$ and TiCl$_3$, the exchange interaction constants are significantly larger than those in CrCl$_3$, resulting in higher-energy magnetic excitations. The maximum excitation energy reaches approximately 241~meV at the zone center ($\Gamma$) in VCl$_3$ and 190~meV at the $M$ point in TiCl$_3$. Despite the differences in energy scales, the magnon dispersions exhibit similar characteristics, particularly around the high symmetry direction $K$, where all three compounds display distinctive topological features.

\subsection{Doped  trichloride materials}

\subsubsection{Defect formation}

The pristine CeCl$_3$ crystallizes in a hexagonal structure with space group $P6_3/m$ and a calculated lattice parameter of 7.43~{\AA}, consistent with experimental value of 7.44~\AA~\cite{zachariasen1948crystal}. The electronic and magnetic properties of Ce-doped CrCl$_3$ are investigated by calculating the electronic band structure and DOS using  a 2 $\times$ 1 $\times$ 1 CrCl$_3$ supercell with a 25\% doping concentration.
The chemical stability of Ce-doped CrCl$_3$ material is confirmed by calculating the defect formation energy under extreme Cl-rich (Cr-poor), Cl-poor (Cr-rich), and Cl-poor (Ce-rich) conditions. In Cl-rich condition, the chemical potential of Cl ($\mu_\text{Cl}$) is taken as bulk energy of a single atom~\cite{limbu2025stability}, i.e., $\mu^\text{{bulk}}_\text{Cl} = E^\text{bulk}_\text{Cl}$. The chemical potentials of Cr ($\mu_\text{Cr}$) and Ce ($\mu_\text{Ce}$) are determined from formation energies of pristine CrCl$_3$ and CeCl$_3$, respectively, using expressions: $\mu_\text{Cr} = E_f(\text{CrCl}_3) + E^\text{bulk}_\text{Cr}$ and $\mu_\text{Ce} = E_f(\text{CeCl}_3) + E^\text{bulk}_\text{Ce}$ under Cl-rich condition. 
For neutral charge state ($q=0$), the defect formation energy is calculated using formula, $E^f_{d} = E_\text{defect} - E_\text{pristine} + \mu_\text{Cr} - \mu_\text{Ce}$, where $E_\text{defect}$ and $E_\text{pristine}$ represent the total ground state energies of Ce-doped and pristine supercells.  
Under Cl-rich condition, the defect formation energy is found to be 0.27~eV. Similarly, for Cl-poor condition, the chemical potential of Cr and Ce are taken as their bulk energies: $\mu_\text{Cr}$ = $E^\text{bulk}_\text{Cr}$ 
and $\mu_\text{Ce}$ =  $E^\text{bulk}_\text{Ce}$. 
Using these chemical potentials, the defect formation energy, under Cl-poor condition, is -4.05~eV. Under Cr-poor and Ce-rich conditions in a Cl-poor environment, the defect formation energy is -8.65~eV, exhibiting the most preferable condition for Ce doping in CrCl$_3$. 

Moreover, at 12.5\% Ce doping, the defect formation energy is found to be -8.52~eV under the same Cr-poor and Ce-rich (Cl-poor) conditions as for the 25\% Ce doping case.  Similarly, we have also computed the defect formation energy of Cr doped VCl$_3$ and found small positive values of 0.06~eV under Cl-rich and 0.53~eV under Cl-poor conditions, indicating that extra energy is required for its formation. However, it becomes -4.53~eV under V-and Cl-poor and Cr-rich conditions, exhibiting the most suitable condition for Cr doped VCl$_3$. Additionally, for Ti doped VCl$_3$, the defect formation energies are -0.25~eV for Cl-rich, -1.58~eV for Cl-poor, and -6.64~eV for V- and Cl-poor and Ti-rich limits, demonstrating its chemical stability. Moreover, defect formation energy for Ce doping on VCl$_3$ (2 $\times$ 2 $\times$ 1) is investigated systematically. The most favorable condition for doping is identified as V- and Cl-poor and Ce-rich, with a defect formation energy of -8.75~eV. Under Cl-poor and Cl-rich conditions, the defect formation energies are 0.18~eV and -3.69~eV, respectively. Therefore, the defect formation energies of Ce-, Cr-, and Ti-doped \(\text{VCl}_{3}\) vary across different chemical potential limits, showing favorable negative values under metal-rich and Cl-poor conditions that indicate stability and feasibility for doping.

\subsubsection{Phonons and magnon-phonon coupling}

\begin{figure}
\centering
\includegraphics[width=0.38\textwidth]{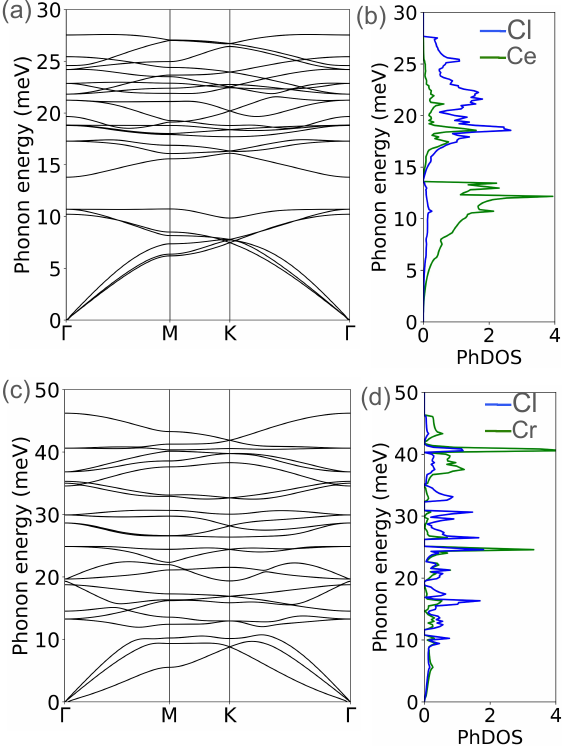}
\caption{Phonon dispersions and phonon density of states, PhDOS (in states/eV atom) of CeCl$_3$ (a, b) and CrCl$_3$ (c, d). The positive phonon frequencies confirm the dynamical stability. An exotic topological
feature is seen at the high symmetry direction $K$.}
\label{phonon}
\end{figure}

In addition to calculating the defect formation energy of Ce-doped CrCl$_3$, we have also calculated the phonon dispersion of pristine CeCl$_3$, obtained by replacing Cr with Ce in CrCl$_3$. Phonon calculations for the Ce-doped CrCl$_3$ material require a sufficiently large supercell to accurately capture the lattice dynamics, making these calculations computationally expensive. Therefore, we use the phonon dispersions of pristine CeCl$_3$ and CrCl$_3$ as a reference to investigate the lattice dynamical stability of the Ce-doped CrCl$_3$ material (Fig.~\ref{phonon}). The positive phonon frequencies confirm the dynamical stability. The lower-frequency phonon modes (below 15~meV) are primarily dominated by the vibrations of Ce atoms in CeCl$_3$, owing to their relatively larger atomic mass as compared with Cl (Fig.~\ref{phonon}b). In contrast, the higher-frequency phonon modes are predominantly associated with the vibrations of Cl atoms. In CrCl$_3$, however, the phonon modes contributed by Cr and Cl are mixed and spanned over the whole energy range because of their similar atomic masses. At the high-symmetry point $K$ in the Brillouin zone, crystal symmetry protects acoustic and low-energy optical phonon branches from splitting, forcing them to cross and form exotic topological features like Dirac points, quadratic touchings, or nodal lines. These band crossings may act as robust highways for vibrational energy, fundamentally altering how low-temperature heat flows through the lattice and how atomic vibrations mix. Interestingly, for CrCl$_3$, there are two such low frequency topological features, namely at $\sim$ 9 meV and $\sim$ 13 meV. The latter matches with the magnon band crossings at $\sim$ 13 meV, thereby indicating a potential magnon-phonon coupling.

\subsubsection{Electronic structure, band gaps, and magnetic properties}

\begin{figure}
\centering
\includegraphics[width=0.47\textwidth]{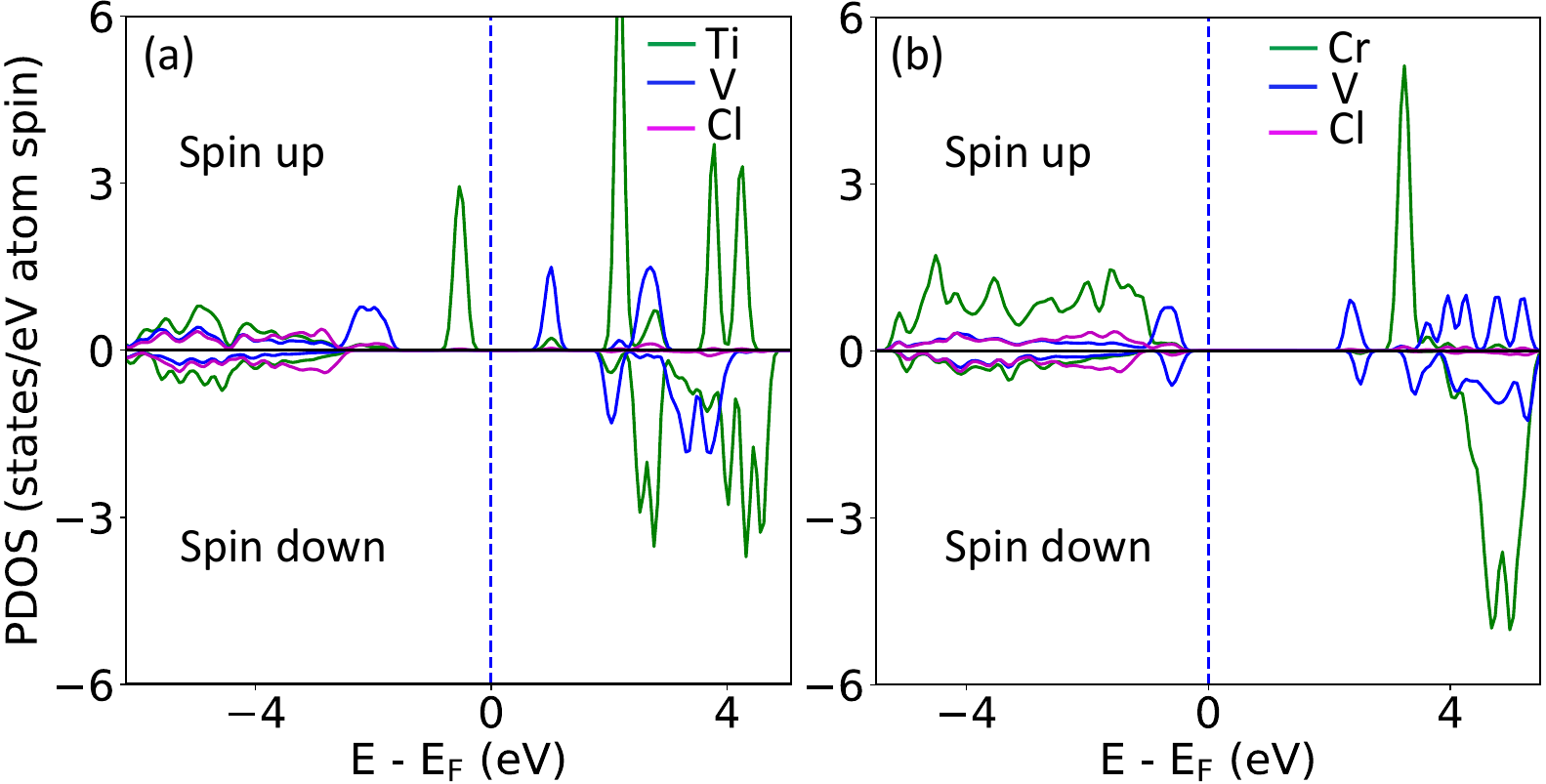}
\caption{PDOS of Ti- (a) and Cr- (b) doped VCl$_3$ calculated using HSE06.}
\label{doping-pdos}
\end{figure}

The electronic and magnetic properties of TM (Cr and Ti) doped VCl$_3$ are also investigated.  Specifically, band structure, DOS, and  magnetic moments are calculated and analyzed using  12.5\% doping concentration with GGA + $U$ ($U$ = 3~eV for both Cr and Ti) approach (Fig.~\ref{doping-pdos}). Cr doped VCl$_3$ is found to be a direct band gap semiconductor with a band gap of 2.59~eV, which is larger than that of the pristine VCl$_3$. Further, the HSE06 calculations yield a band gap of 2.74~eV, which is larger than that found using GGA + $U$ approach. In contrast, Ti doping reduces the band gap to 1.29~eV (1.32~eV) with GGA + $U$ (HSE06), indicating that dopants with a single valence-electron difference can significantly alter the electronic properties. These results also suggest that the band gap of VCl$_3$ can be effectively tuned via TM doping. For Ti-doped VCl$_3$, the bottom of the conduction band, in both spin up and spin down channels, is contributed by V-$3d$ state with negligible contribution from Cl-$2p$ and Ti-$3d$ states (Fig.~\ref{doping-pdos}(a)). Interestingly, a sharp Ti-$3d$ state is identified just below the Fermi level, producing a spin magnetic moment of 0.92~$\mu_B$, while having a negligible effect on the spin magnetic moment of V (1.92~$\mu_B$). In contrast to Ti doping, the conduction and valence band edges in the spin-up and spin-down channels are mainly governed by V-$3d$ states in Cr-doped VCl$_3$ (Fig.~\ref{doping-pdos}(b)), with minimal hybridization with Cl-$2p$ states. In the conduction band, the Cr dopant contributes asymmetrically to the PDOS and exhibits a spin magnetic moment of 2.934~$\mu_B$.

\begin{figure}
\centering
\includegraphics[width=0.47\textwidth]{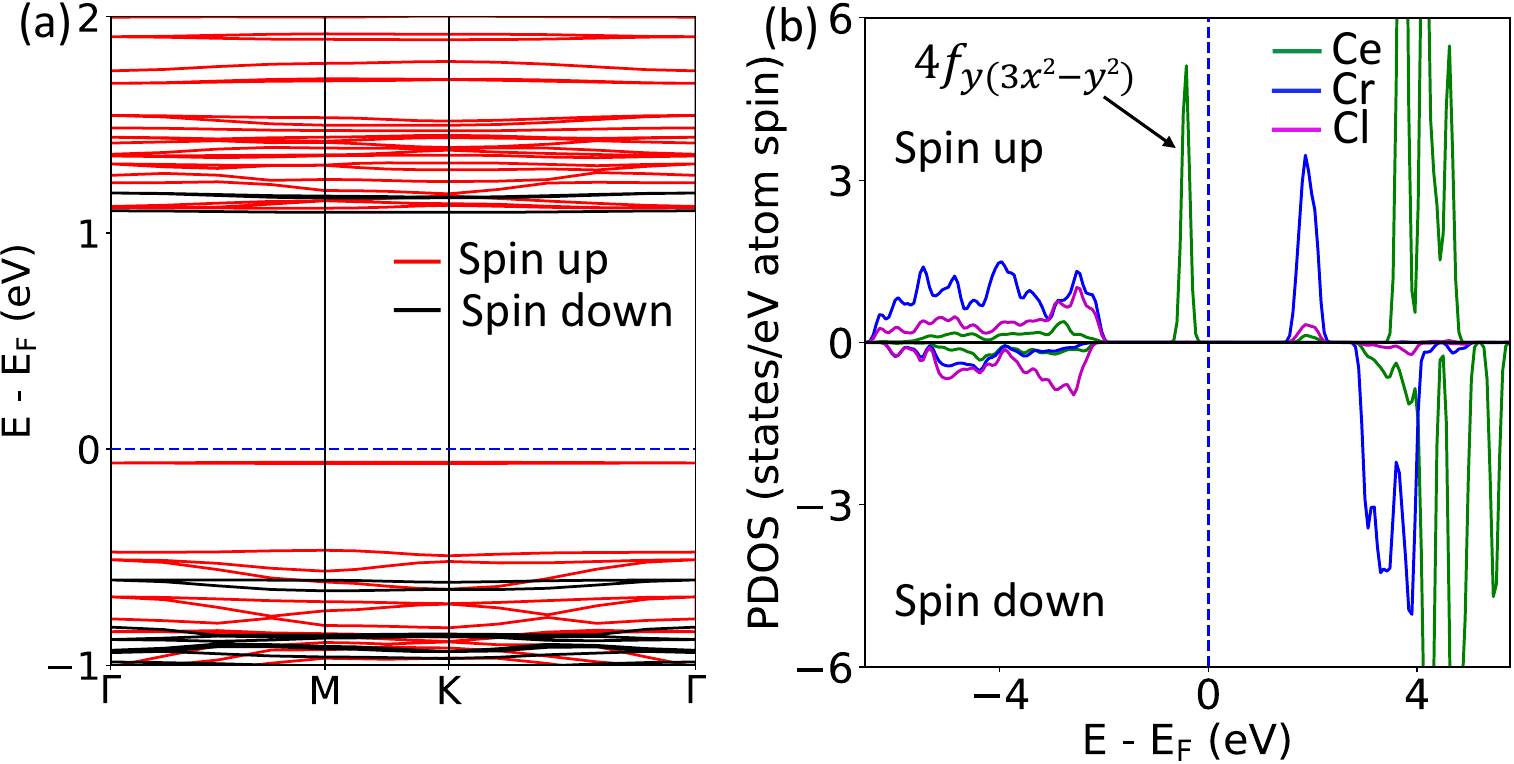}
\caption{Band structure (a) and corresponding PDOS per atom (b) of Ce-doped CrCl$_3$ calculated using the HSE06.}
\label{Ce-dopin}
\end{figure}

The electronic band structure of pristine and Ce-doped CrCl$_3$ exhibit direct band gaps of 2.42~eV and 0.20~eV, respectively, in the majority spin channel, as obtained from GGA + $U$ calculations with $U$ = 3~eV for Cr~\cite{limbu2025magnetic} and $U$ = 5~eV for Ce~\cite{loschen2007first}. A sharp localized $4f$ state contributed solely from $f_{y(3x^2-y^2)}$ is appeared just below the Fermi level, producing a spin magnetic moment of 0.89~$\mu_B$ per Ce. The $4f$ electron of Ce atom is predominantly localized in the $f_{y(3x^2-y^2)}$ orbital, indicating a well-defined occupied quantum state. HSEO6 also produces a similar PDOS feature with band gap of 1.16~eV, showing a spin magnetic moment of 0.98~$\mu_B$ per Ce (Fig.~\ref{Ce-dopin}). Above the Fermi level, in bottom of the conduction band, there is significant contribution by Cr atom with minor contribution from  Cl and Cr atoms. Similarly, Ce-doped TiCl$_3$ and VCl$_3$ also exhibit FM semiconductors, 
exhibiting a spin magnetic moment of $\sim$ 1~$\mu_B$ per Ce. Additionally, the calculated orbital magnetic moment of Ce (Cr) is 0.52~$\mu_B$ (0.03~$\mu_B$) from GGA + $U$ + SOC calculations, indicating substantial quenching by the crystalline electric field arising from the surrounding atoms. 

In addition, we investigated the influence of TM (Cr and Ti) and Ce dopants on magnetic properties of VCl$_3$ and CrCl$_3$, respectively. The ground state magnetism of doped material is determined from the total energy comparison of FM ($E_\text{FM}$) and AFM ($E_\text{AFM}$) spin configurations~\cite{limbu2025stability}, $\Delta E = E_\text{AFM} - E_\text{FM}$. For Cr and Ti doped VCl$_3$, the total energies of the FM configurations are lower than those of the AFM configurations by 28.78~meV and 0.20~meV, respectively, confirming the FM ground state. Similarly, Ce doped CrCl$_3$ also energetically prefers FM spin configuration with energy 0.091~eV lower than AFM spin configuration. In addition, two Ce atoms are substituted at first-nearest-neighbor Cr sites, corresponding to a doping concentration of 11.11\% (3 $\times$ 3 $\times$ 1 supercell). The magnetic exchange interaction between the Ce atoms is calculated to be 35.04~meV, indicating a FM configuration. In pristine 2 $\times $ 2 $\times$ 1 supercell of VCl$_3$, the total magnetic moment is 16~$\mu_B$ with  a spin magnetic moment of 1.83~$\mu_B$ per V. However, the total magnetic moment becomes 16.99~$\mu_B$ for Cr doped VCl$_3$ with spin magnetic moment of 2.83~$\mu_B$ for Cr, without changing the magnetic moment of V. Similarly for Ti doped VCl$_3$, we find the total magnetic moment of 14.99~$\mu_B$ per supercell with a spin magnetic moment of 0.12~$\mu_B$ for Ti. In addition to the TM doped material, Ce doped CrCl$_3$ also preserves a FM character with total magnetic moment of 10~$\mu_B$ per supercell with 0.89~$\mu_B$ for Ce and 2.86~$\mu_B$ for Cr.

\section{Conclusion}

Utilizing advanced DFT calculations, we have studied  phase stability, electronic structure, magnetic interactions, phononics and magnonics of selected pristine and $3d$-electrons (Ti and Cr) doped VCl$_3$, and $4f$-electron (Ce) doped CrCl$_3$. The negative cohesive and formation energies confirm the stability of the pristine phases. Meanwhile, negative defect formation energies show that the doped phases are chemically stable and favored in dopant-rich conditions. The positive phonon frequencies in cerium trichloride prove that adding cerium to transition metal trichlorides creates a stable crystal structure. The phonon dispersions in this
material exhibits exotic topological feature at a special symmetry direction $K$.  The strongly correlated $3d$-electrons of transition metals are treated using the HSE06 approach. Consistent with available literature, these hybrid HSE06 calculations produce FM semiconducting behavior with band gaps of 2.54~eV for VCl$_3$, 4.01~eV for CrCl$_3$, and 2.61~eV for TiCl$_3$. The positive values of the calculated magnetic exchange interactions also confirm the FM behavior in VCl$_3$, CrCl$_3$, and TiCl$_3$. Using these calculated exchange constants, the magnon dispersions are obtained by diagonalizing the spin Hamiltonian within the Holstein-Primakoff formalism. The resulting dispersions show two magnon branches that are associated with the two magnetic sublattices. The two magnon branches cross at the high symmetry $K$ point, indicating a topological feature. This topological magnonic feature at $K$ correlates with the topological feature seen in phonons leading to magnon-phonon coupling. The calculated magnon dispersion of CrCl$_3$ matches the energy scale and behavior of the experimentally observed magnon spectrum in CrI$_3$. Doping VCl$_3$ with chromium increases the band gap to 2.74 eV, while doping with titanium reduces it to 1.32 eV. Both methods keep the material FM. In addition, Ce-doped CrCl$_3$ remains a FM semiconductor, yielding a band gap of 1.16~eV, and exhibiting a localized $4f$ state just below the Fermi level, with a spin magnetic moment of approximately 1 $\mu_B$. When two Ce atoms are doped at first-nearest-neighbor Cr sites, the material energetically favors a FM configuration. This state is accompanied by a large magnetic exchange constant, indicating the potential for a robust, high-temperature FM semiconductor. Therefore, the tunability of the electronic and magnetic properties of these magnetic semiconductors via $3d$-and $4f$-electrons doping underscores their significant promise in defect-engineered spintronics. \\\\\\

\section*{Acknowledgments}

\noindent This work was supported as part of the Center for Energy Efficient Magnonics, an Energy Frontier Research Center funded by the U.S.\@ Department of Energy, Office of Science, Basic Energy Sciences, under Award number DE-AC02-76SF00515. KPC is supported by University Grants Commissions (UGC) Nepal by providing PhD fellowship under Award number Ph.D. 80/81-S \& T- 11. DP acknowledges the use of the computational facilities on the Frontera supercomputer at the Texas Advanced Computing Center (TACC) via the pathway allocation, DMR23051.

\bibliography{references}

\end{document}